\documentclass[runningheads]{llncs}
\usepackage[dvipsnames]{xcolor}
\definecolor{Su}{RGB}{76,114,176}
\definecolor{In}{RGB}{196,78,82}
\definecolor{Re}{RGB}{85,168,104}

\usepackage[utf8]{inputenc}
\usepackage{amsmath}
\usepackage{amsfonts}
\usepackage{graphicx}
\usepackage{color}
\usepackage{subfig}

\usepackage{tikz}
\usetikzlibrary{positioning,calc}
\usepackage{color}

\usepackage[misc]{ifsym} 

\usepackage{mathtools}

\usepackage{algorithm}
\usepackage[noend]{algpseudocode}

\makeatletter
\def\algbackskip{\hskip-\ALG@thistlm}
\makeatother

\begin{document}
	\title{Rejection-Based Simulation of Stochastic Spreading Processes on Complex Networks}
	\titlerunning{Rejection-Based Simulation of Stochastic Processes on Networks}
	%
	\author{Gerrit Großmann\textsuperscript{(\Letter)} \orcidID{0000-0002-4933-447X} \and
		Verena Wolf}
	\authorrunning{Großmann et al.}
	%
	\institute{Saarland University, 66123 Saarbrücken, Germany \\
		\url{mosi.cs.uni-saarland.de}  \\
		\email{\{gerrit.grossmann,verena.wolf\}@uni-saarland.de} }
	\maketitle              
	\begin{abstract}
	    Stochastic processes can model many emerging phenomena on networks, like the spread of computer viruses, rumors, or infectious diseases. 
		Understanding the dynamics of such stochastic spreading processes is therefore of fundamental interest.
		In this work we consider the wide-spread compartment model where each node is in one of several states (or compartments).
		Nodes change their state randomly after an exponentially distributed waiting time and according to a given set of rules.
		For networks of realistic size, even the generation of only a single stochastic trajectory of a spreading process  is computationally very expensive. 
		
		Here, we propose a novel simulation approach, which combines the advantages of event-based simulation and rejection sampling.
		Our method outperforms state-of-the-art methods  in terms of absolute runtime and scales significantly better, while being statistically equivalent.  
		
	\end{abstract}

	\section{Introduction}
	\raggedbottom
	Computational modeling of spreading phenomena is an active research field within network science with many applications ranging from disease prevention to social network analysis  \cite{barabasi2016network,barrat2008dynamical,porter2016dynamical,goutsias2013markovian,pastor2015epidemic,kiss2016mathematics}. 
	The most widely used approach is a continuous-time model where each node of a given graph occupies one of several states (e.g.\;\emph{infected} and \emph{susceptible}) at each point in time. 
	A set of rules determines the probabilities and random times at which nodes change their state depending on the node's direct neighborhood (as determined by the graph). 
	The application of a rule is always stochastic and the waiting time before a rule ``fires'' (i.e.\;is applied) is governed by an exponential distribution. 
	
	The underlying stochastic dynamics are given by a continuous-time Markov chain (CTMC) \cite{kiss2016mathematics,simon2011exact,van2009virus,sahneh2013generalized}. 
	Each possible assignment from nodes to local node states constitutes an individual state of the CTMC (here referred to as \emph{CTMC state} or \emph{network state} to avoid confusion with the local state of a single node). 
	Hence, the corresponding CTMC state space grows exponentially in the number of nodes, which renders its numerical solution infeasible. 
	
	As a consequence, mean-field-type approximations and sampling approaches have emerged as the cornerstones for their analysis. Mean-field equations originate from statistical physics and provide typically reasonable good approximation of the underlying dynamics 
	\cite{gleeson2011high,gleeson2012accuracy,gleeson2013binary,devriendt2017unified,bortolussi2013continuous}.
	Generally speaking, they propose a set of ordinary differential equations that models the average behavior of each component (e.g., for each node, or for all nodes of certain degree). 
	However, mean-field approaches only give information about the \emph{average behavior} of the system, for example, about the expected number of \emph{infected} nodes for each degree.
	Naturally, this restricts the scope of their application. In particular, they are not suited to answer specific questions about the system.
	
	For example, one might be interested in finding the specific source of an epidemic \cite{prakash2012spotting,farajtabar2015back} or wants to know where an intervention (e.g.\;by vaccination) is most successful. \cite{schneider2011suppressing,cohen2003efficient,buono2015immunization,wu2015influence}. 
	
	
	Consequently, stochastic simulations remain an essential tool in the computational analysis of complex networks dynamics. {State-of-the-art methods will be revised in Chapter \ref{literature}.} 

	Here, we propose an event-driven simulation method which utilizes rejection sampling. 
	Our method is based on a event queue which stores infection and curing events.
	Unlike traditional methods, we ensure that it is not necessary to iterate over the entire neighborhood of a node after it has changed its state. 
	Therefore, we allow the creation of events which are inconsistent with the current CMTC state. These might lead to rejections when they reach the beginning of the queue.
	We introduce our method for the well-known SIS (Susceptible-Infected-Susceptible) model and show that it can easily be generalized for other epidemic-type processes. 
	Code will be made available.\footnote{\url{github.com/gerritgr/Rejection-Based-Epidemic-Simulation}}
	
	We formalize the semantics of spreading processes in Section \ref{semantics} and explain how the CTMC is constructed.
	To sample the CTMC, different statistically equivalent algorithms exist, which we present in Section \ref{literature}. In Section \ref{ourmethod} we present our rejection sampling algorithm. We demonstrate the effectiveness of our approach on three different case studies in Section \ref{casestudies}. 
	
	\section{Stochastic Spreading Processes\label{semantics}}
	Let $\mathcal{G}=(\mathcal{N}, \mathcal{E})$ be a an undirected, unweighted, finite graph without self-loops. We assume the edges are tuples of nodes and that $(n_1,n_2) \in \mathcal{E}$ always implies $(n_2,n_1) \in \mathcal{E}$.
	At each time point $t \in \mathbb{R}_{\geq 0}$ each node occupies one out of $m$ (local) states (also called labels or compartments), denoted by $\mathcal{S} = \{s_1, s_2, \dots, s_m\}$.
	Consequently, the (global) network state is fully specified by a labeling $L: \mathcal{N} \rightarrow \mathcal{S}$.
	We use $\mathcal{L} = \{L \mid L: \mathcal{N} \rightarrow \mathcal{S} \}$ to denote all possible network states. As each of the $|\mathcal{N}|$ nodes occupies one of $m$ states, we know that $|\mathcal{L}| = m^{|\mathcal{N}|}$.
	Nodes change their state by the application of a stochastic rule. 
	A node's state and its neighborhood determine which rules are applicable to a node and the probability density of the random delay until a rule fires. 
	If several rules can fire, the one with the shortest delay is executed.  
	
	We allow two types of rules: \emph{node-based (independent, spontaneous)} rules and \emph{edge-based (contact, spreading)} rules.
	The application  of a node-based rule $\text{A} \xrightarrow{\mu} \text{B}$ results in a transition of a node from state $\text{A} \in \mathcal{S}$ to state $\text{B} \in \mathcal{S}$  ($A \neq B$) with rate $\mu \in \mathbb{R}_{> 0}$. 
	That is, the waiting time until the rule fires is governed by the exponential distribution with rate $\mu$.  
	An edge-based rule has the form $\text{A}+\text{C} \xrightarrow{\lambda} \text{B}+\text{C}$, where $\text{A},\text{B},\text{C} \in \mathcal{S}, \text{A} \neq \text{B}, \lambda \in \mathbb{R}_{> 0}$.
	Its application changes an edge (more precisely, the state of an edge's node).  
	It can be applied to each edge $(n,n') \in \mathcal{E}$ where $L(n)=\text{A}, L(n')=\text{B}$. 
	Again, the node in state $\text{A}$ changes after a delay that is exponentially distributed with rate $\lambda$.  
	Note that, if a node in state $\text{A}$ has more than one direct $\text{B}$-neighbor, it is ``attacked'' independently by each neighbor. 
	Due to the properties of the exponential distribution, the rate at which a node changes its state according to a certain contact rule is proportional to the number of neighbors which induce the change. 
	
	\subsubsection{SIS Model}
	In the sequel, we use the well-known Susceptible-Infected-Susceptible (SIS) model as a running example. Consider $\mathcal{S} = \{\text{I}, \text{S}\}$ and the rules:
	
	\begin{equation*}
	\text{S}+\text{I}  \xrightarrow{\lambda} \text{I}+\text{I}  \hspace{1.0cm}
	\text{I}  \xrightarrow{\mu} \text{S}\;\;.
	\end{equation*}
	
	In the SIS model, infected nodes propagate their infection to neighboring susceptible nodes using an edge-based rule. 
    Thus, only susceptible nodes with at least one infected neighbor can become infected. Infection of a node occurs at a rate that increases proportionally with the number of infected neighbors.
	Infected nodes can, independently from their neighborhood, recover (i.e.\;become susceptible again) using a node-based rule.  
	
	\section{Previous Approaches\label{literature}}
	In this section, we shortly revise techniques that have been previously suggested for the simulation of SIS-type processes. For a more comprehensive description we refer the reader to \cite{kiss2016mathematics,cota2017optimized}.

	\subsection{Standard Gillespie Algorithm}
	The Standard Gillespie Algorithm (here, simply referred to as \emph{GA}) is  also known as Gillespie’s direct method and a popular method for the simulation of coupled chemical reactions. Its adaptation to complex networks uses  as key data structures
	 two lists  which are constantly updated: a list of all infected nodes (denoted by $\mathcal{L}_{I}$) and a list of all S--I edges (denoted by $\mathcal{L}_{S-I}$). 
		
		In each simulation step, we first draw an exponentially distributed delay for the time until the next rule fires.  
	That is, instead of sampling a waiting time for each rule and each position where the rule can be applied, we directly sample the time until the network state changes. 
	For this, we compute an aggregated rate $c = \mu  |\mathcal{L}_{I}| + \lambda   |\mathcal{L}_{S-I}| $.
	Then we randomly decide if an infection or a curing event is happening.
	The probability of the latter is proportional to its rate, i.e.\;$\frac{1}{c} \mu |\mathcal{L}_{I}| $, and thus, the probability of an infection  is 
	$\frac{1}{c} \lambda |\mathcal{L}_{S-I}|$.
	After that we pick an infected node (in case of a curing) or an S--I edge (in case of an infection) uniformly at random. We update the two lists accordingly.
	The expensive part in each step is keeping $\mathcal{L}_{S-I}$ updated. 
	For this, we iterate over the whole neighborhood of the node and for each susceptible neighbor we remove (after a curing) or add (after an infection) the corresponding edge to the list. Thus, we need one add/remove operation on the list for each susceptible neighbor. 
	
	Note that are different possibilities to sample the node that will become infected next. 
	Instead of keeping an updated list of all S--I edges one can also use a list of all susceptible nodes. In that case, we cannot sample uniformly but decide for the infection of a susceptible node with a probability proportional to its number of infected neighbors. 
	
	Likewise, we can randomly pick the starting point of the next infection by only considering $\mathcal{L}_{I}$.
	To generate an infection event, we first sample an infected node from this list and then we (uniformly) sample a susceptible neighbor, which becomes infected. 
	Since infected nodes with many susceptible neighbors have a higher probability of being the starting point of an infection (i.e., they have more S--I edges associated with them), we sample from $\mathcal{L}_{I}$ such that the probability of picking an infected node is proportional to its number of susceptible neighbors.

	All three approaches are statistically equivalent but the last one motivates the \emph{Optimized Gillespie Algorithm} (OGA)   \cite{cota2017optimized}.
	
	\subsection{Optimized Gillespie Algorithm}

	As discussed earlier, 
	sampling from $\mathcal{L}_I$ is expensive because 
	for each infected node
	it is necessary to store its corresponding number of susceptible neighbors. 
	Updating this information for all elements of $\mathcal{L}_I$
	is costly
	because after each event, the number of susceptible neighbors may change for many nodes.

\begin{figure}[t]
		\centering
		\includegraphics[width=0.3\linewidth]{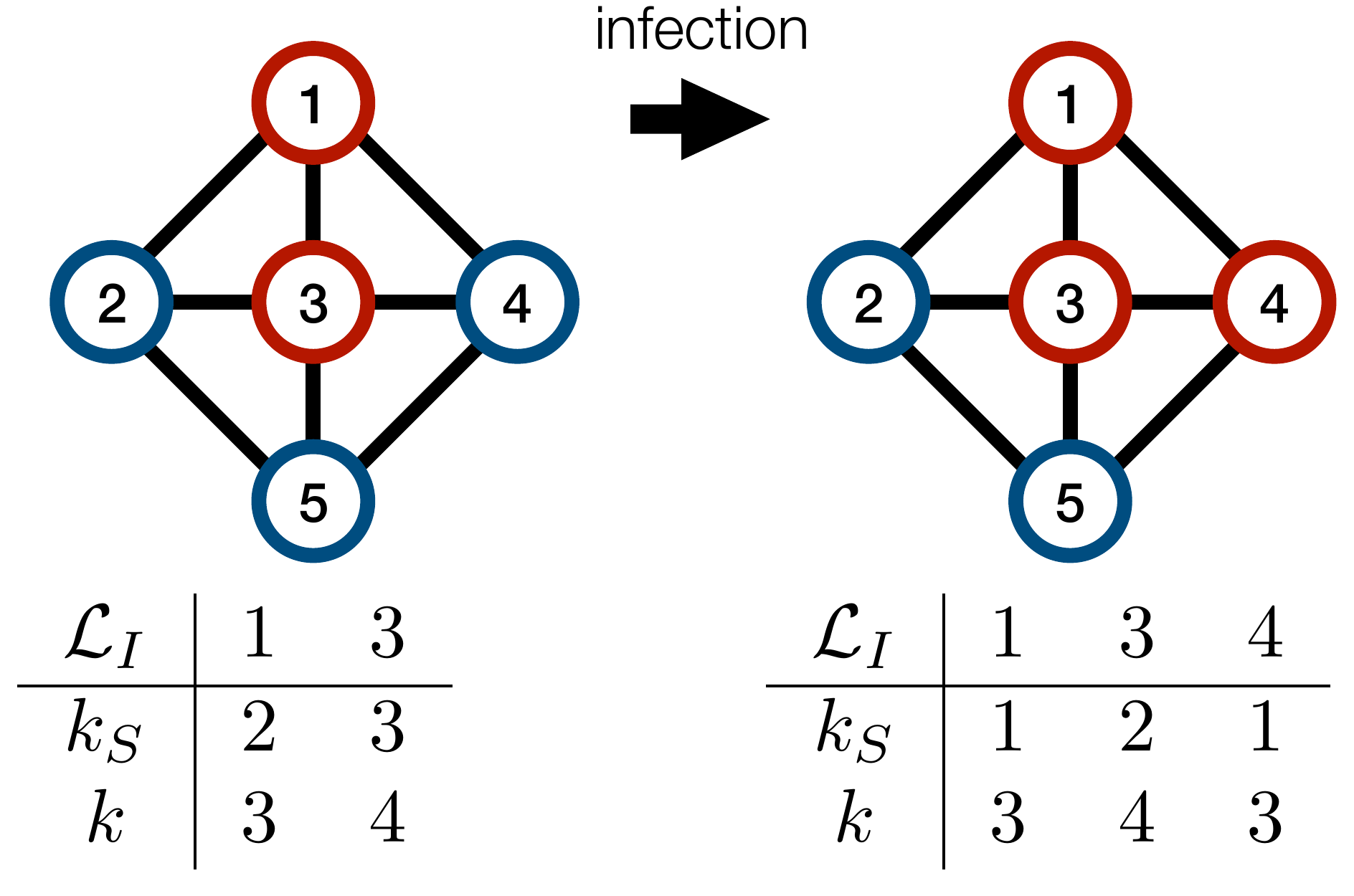}
		\caption[Infection event rejection]{Example of an infection event. 
		We sample from $\mathcal{L}_I$ proportional to $k_S$.  Alternatively, we can weight according to the number of neighbors $k$ which is constant and over-approximates $k_S$.  To correct for the over-approximation we reject a sample with probability $\frac{k-k_s}{k}$. }
		\label{fig:infectionsamplecut}
	\end{figure}

	  Cota and Ferreira \cite{cota2017optimized} suggest to sample nodes from $\mathcal{L}_I$  with a probability that is proportionally to the degree $k$ of a node, which is an upper bound for the maximal possible number of susceptible neighbors. Then they
	  uniformly choose a neighbor of that node and update the global clock.
	If this neighbor is already infected they reject the infection event, which 
	  yields a rejection probability of $\frac{k-k_S}{k}$ if $k_S$ is the 
	  number of susceptible neighbors.
	Note that the rejection probability exactly corrects for the over-approximation of using $k$ instead of $k_S$. This is illustrated in
	Fig.~\ref{fig:infectionsamplecut}.
		
	Compared to the GA, updating the list of infected nodes becomes cheaper, because only the node which actually changes its state is added to (or removed from) $\mathcal{L}_I$. The sampling probabilities of the neighbors remain the same because their degree remains the same. 
	On the other hand, sampling of a node is more expensive compared to the GA where we sample edges uniformly. 
	
	Naturally, the speedup in each step comes at the costs (of a potentially enormous amount) of rejection events. 
	Even a single infected node with many infected but few susceptible neighbors will continuously lead to rejected events. This is especially problematic in cases with many infected nodes and no or very few susceptible neighbors which therefore make rejections many orders of magnitude more likely than actual events. 
	Therefore, in \cite{cota2017optimized} the authors propose the algorithm for simulations close to the epidemic threshold, where the number of infected nodes is typically very small. 
	
	\subsection{Event-Based Simulation} \label{subsec_event_based}
	In the event-driven approach 
	the primary data structure is an event queue, in which events are sorted and executed according to the time points at which they will occur. 
	This eliminates the costly process of randomly selecting a node for each step (popping the first element from the queue has constant time complexity). Events are either curing of a specific node or infection via a specific edge. {Moreover, it is easy to adapt the event-driven approach to rules with non-Markovian waiting times or to a network where each node has individual recovery and infection rates \cite{kiss2016mathematics}.
	Event-based simulation of an SIS process is done as follows:
	For the initialization, we draw for each node an exponentially distributed  time until recovery with rate $\mu$ and add the respective curing event for the node to the queue. 
	Likewise, for each susceptible node with at least one infected neighbor we draw an exponentially distributed  time until infection with rate $\lambda \times \text{``Number of infected neighbors''}$. We add the resulting events to the queue. 
	
	During the simulation, we always take the earliest event from the queue, 
	change  the network accordingly and update the global clock. 
	If the current event is the infection of a node, the infection rates of its susceptible neighbors increase. 
	Likewise, if the current event is a recovery of a node the infection rates of its susceptible neighbors decrease. 
	Thus, it is necessary to iterate over all neighbors of the corresponding node, draw renewed waiting times for their infection events, and update the event queue accordingly. 
	Although, efficient strategies have been suggested \cite{kiss2016mathematics}, these queue updates are rather costly.
	
	Since each step requires an iteration over all neighbors of the node under consideration, the wost-case runtime depends on the maximal degree of the network. 
	Moreover, for each neighbor, it might be necessary to reorder the event queue. 
	The time complexity of reordering the queue depends (typically logarithmically) on the number of elements in the queue and adds significant additional costs to each step. 
	Note that trajectories generated using the event-driven approach are statistically equivalent to those generated with the GA, because all
	delays are exponentially distributed and thus have the memoryless property. A variant of this algorithm can also be found in \cite{sahneh2017gemfsim}.
	
	\section{Our Method\label{ourmethod}}
	In this section, we propose a method for the simulation of SIS-type processes. 
	The key idea is to combine an event-driven approach with rejection sampling while keeping the number of rejections to a minimum.
	We will generalize the algorithm for different epidemic processes as well as for weighted and temporal networks. 

	 First, we first introduce the main data structures. 
	\paragraph{Event queue}
		It stores all future infection and curing events generated so far. 
		Each event is associated with a time point and with the node(s) affected by the event. Curing events contain a reference to the recovering node and infection events to a pair of connected nodes, an infected (source) node and a susceptible (target) node. 
		\paragraph{Graph} In this graph structure, each node is associated with its list of neighbors, its current state, a degree, and, if infected, a prospective recovery time.  \\
		
	We also keep track of the time in a global clock.
	We assume that an initial network, a time horizon (or another stopping criterion), and the rate parameters ($\mu, \lambda$) are given as input. 
	In Alg. \ref{init}-\ref{mainproc} we provide pseudocode for the detailed steps of the method.
	
	\subsubsection{Initialization}    
	Initially, we iterate over the network and sample a recovery time (exponentially distributed with rate $\mu$) for each infected node 
	(cf. Line 2, Alg. \ref{init}).
	We push the recovery event to the queue and annotate each infected node with its recovery time (cf. Line 5, Alg. \ref{gen_rec}).
	Next, we iterate over the network a second time and generate an infection event for each infected node as explained later (cf. Line 5, Alg. \ref{init}). 
	We need two iterations because the recovery time of each infected node has to be available for the infection events. 
	This event identifies the earliest infection attempt of the current node which, according to the current graph, has a certain chance of success.
	The generation of infection events and the distinction between unsuccessful and potentially successful infection attempts is an essential part of the algorithm.
 	
	\subsubsection{Iteration} 
	The main procedure of the simulation is illustrated in Alg. \ref{mainproc}.
	We schedule events until the global clock reaches the specified time horizon
	(cf. Line 9). 
	In each step, we take the earliest event from the   queue (Line 7) and set the global clock to the event time (Line 8). Then we ``apply'' the event (Line 11-20). 
	
	In case of a recovery event, we simply change the state of the corresponding node from I to S and are done (Line 12). 
	Note that we always generate (exactly) one recovery event for each infected node, thus, each recovery event is always consistent with the current network state. Note that the queue always contains exactly one recovery event for each infected node.
	
	If the event is an infection event, we apply the event if possible (Line 14-18) and reject it otherwise (Line 19-20). 
	We update the global clock either way. Each infection event is associated with a source node and a target node (i.e., the node under attack). 
	The infection event is applicable if the current state of the target node is S  (which might not be the case anymore) and the current state of the source node is I (which will always be the case). 
	After a successful infection event, we generate a new recovery event for the target node (Line 16) and two novel infection events, one for the source node (Line 17) and one for the target node which is now also infected (Line 18). 
	If the infection attempt was rejected, we only generate a novel infection event for the source node (Line 20). 
	
	\subsubsection{Generating Infection Events} 
In Alg. \ref{gen_inf} we describe the generation of infection events.	
	For each infected node we only generate the earliest infection attempt and add it to the queue. Therefore, we first sample the exponentially distributed waiting time with rate $k \lambda$, where $k$ is the degree of the node, and compute the time point of infection (Line 5).
	If the time point of the infection attempt is after its recovery event, we stop and no infection event is added to the queue (Lines 6-7).
	Note that in the graph structure, each node is annotated with its recovery time (denoted as 
	node.recovery$\_$time) to have it immediately available. 
	
	Next, we uniformly select a random neighbor which will be attacked (Line 8). 
	If the neighbor is currently susceptible, we add the event to the event queue and 
	the current iteration step ends (Lines 9-12). 
	
	If the neighbor is currently infected we check the recovery time of the neighbor (Line 9). 
	If the infection attempt happens before the recovery time point, we already know that the infection attempt will be unsuccessful (already infected nodes cannot become infected). 
	Thus, we perform an \emph{early reject} (Lines 10-12 are not executed).
	That is, instead of pushing the surely unsuccessful infection event to the queue, we directly generate another infection attempt, i.e. we re-enter the while-loop in Lines 4-12. 
	We repeat the above procedure until the recovery time of the current node is reached
	or the infection can be added to the queue (i.e.\;no early rejection is happening).
	
	Fig.~\ref{fig:fullpage} provides a minimal example of a potential execution of our method. 

\begin{figure}
\scalebox{0.8}{
\begin{minipage}{1.1\linewidth}
\begin{algorithm}[H]
	\caption{Graph Initialization}\label{init}
	\begin{algorithmic}[1]
		\Procedure{InitGraph}{$G$, $\mu$, $\lambda$, $Q$}
		\For{\textbf{each} node in $G$} 
		\If{node.state = I} 
		\State  \textsc{GenerateRecoveryEvent}(node, $\mu$, 0, $Q$)
		\EndIf
		\EndFor
		\For{\textbf{each} node in $G$} 
			\Comment recovery times are available now
		\If{node.state = I} 
		\State \textsc{GenerateInfectionEvent}(node, $\lambda$, 0, $Q$)
		\EndIf
		\EndFor
		\EndProcedure
	\end{algorithmic}
\end{algorithm}
\vspace{-1.2cm}
\begin{algorithm}[H]
	\caption{Generation of a Recovery Event}\label{gen_rec}
	\begin{algorithmic}[1]
		\Procedure{GenerateRecoveryEvent}{node, $\mu$, $t_{\text{global}}$, $Q$}
		\State $t_{\text{event}}$ = $t_{\text{global}}$ + draw\_exp($\mu$)
		\State e = Event(src\_node = node, t=$t_{\text{event}}$, type=recovery)
		\State node.recovery\_time = $t_{\text{event}}$
		\State $Q$.push(e)
		\EndProcedure
	\end{algorithmic}
\end{algorithm}
\vspace{-1.2cm}
\begin{algorithm}[H]
	\caption{Generation of an Infection Event}\label{gen_inf}
	\begin{algorithmic}[1]
		\Procedure{GenerateInfectionEvent}{node, $\lambda$, $t_{\text{global}}$, $Q$}
		\State  $t_{\text{event}}$ = $t_{\text{global}}$
		\State rate = $\lambda*$node.degree
		\While {true}
		\State $t_{\text{event}}$ +=  draw\_exp(rate)
		\If{node.recovery\_time $<$ $t_{\text{event}}$} 
		\Comment no event is generated
		\State break
		\EndIf
		\State attacked\_node = draw\_uniform(node.neighbor\_list)
		\If{attacked\_node.state = S \newline \hspace*{4em} \textbf{or} attacked\_node.recovery\_time $<$ $t_{\text{event}}$}
		\Comment check for early reject
		\State e = Event(src\_node=node, target=attacked\_node, \newline \hspace*{9em}  time=$t_{\text{event}}$, type=infection)
		\State $Q$.push(e)
		\Comment was successful 
		\State break
		\EndIf
		\EndWhile
		\EndProcedure
	\end{algorithmic}
\end{algorithm}
\vspace{-1.2cm}
\begin{algorithm}[H]
	\caption{SIS Simulation}\label{mainproc}
	\hspace*{\algorithmicindent} \textbf{Input:} Graph ($G$) with initial states, time horizon ($h$), recovery rate ($\mu$), infection rate ($\lambda$) \\
	\hspace*{\algorithmicindent} \textbf{Output:} Graph at time $h$
	\Comment or any other measure of interest
	\begin{algorithmic}[1]
		\State $Q$ = \textsc{emptyQueue}()
			\Comment sorted w.r.t.\;time
		\State \textsc{InitGraph}($G$, $\mu$, $\lambda$, $Q$)
		\State $t_{\text{global}}$ = 0
		\While{true}
		\If{Q.is\_empty()} \State break \EndIf
		\State e = Q.pop()
		\State $t_{\text{global}}$ = e.time
		\If{$t_{\text{global}} > h$} \State break \EndIf
		\If{e.type = recovery} 
		\State G[e.src\_node].state = S 
		\Else 
		\If{G[e.target\_node].state = S}
		\State G[e.target\_node].state = I
		\State \textsc{GenerateRecoveryEvent}(e.target\_node, $\mu$,  $t_{\text{global}}$, $Q$)
		\State \textsc{GenerateInfectionEvent}(e.src\_node, $\lambda$,  $t_{\text{global}}$, $Q$)
		\State \textsc{GenerateInfectionEvent}(e.target\_node, $\lambda$,  $t_{\text{global}}$, $Q$)
		\Else
		\Comment late reject
		\State  \textsc{GenerateInfectionEvent}(e.src\_node, $\lambda$,  $t_{\text{global}}$, $Q$)
		\EndIf
		\EndIf
		\EndWhile
	\end{algorithmic}
\end{algorithm}
\end{minipage}	
}
\caption[Rejection Sampling]{Pseudocode for our event-based rejection sampling method.}
\label{fig:pseudocode}
\end{figure}   
	\begin{figure}
	\centering
	\includegraphics[width=0.8\linewidth]{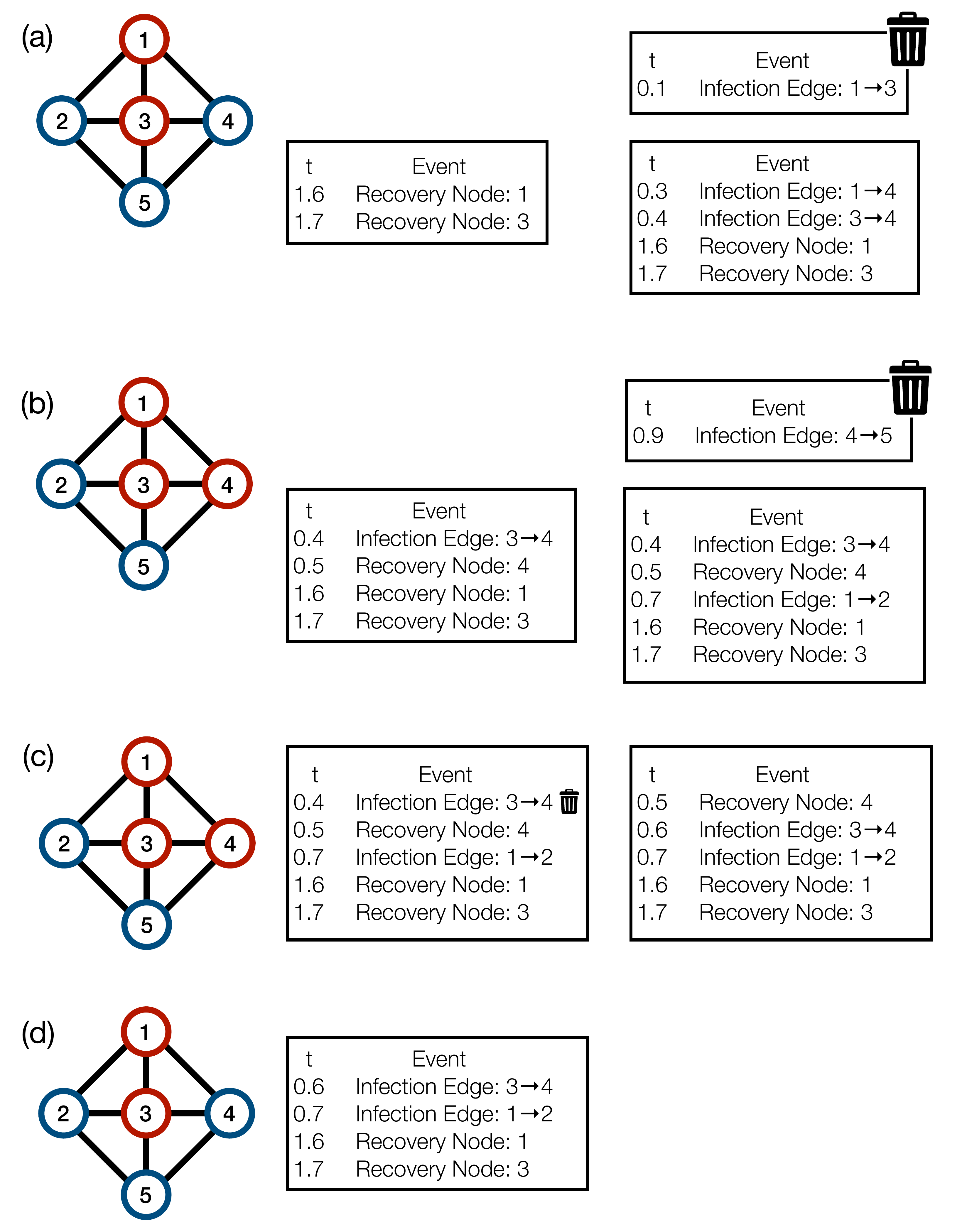}
	\caption[Example Run]{First four steps  of the our method for a toy example (I: red, S: blue): \textbf{(a)} Initialization, generate the recovery events (left queue), and  infection event for each infected node (right queue). The first infection attempt from node $1$ is an early reject. 
		\textbf{(b)} The infection from $1$ to $4$ was successful, we generate a recovery event for $4$ and two new infection events for $1$ and $4$. The infection event of node $4$ is directly rejected because it happens after its recovery.  \textbf{(c)} (Late) Reject of  the infection attempt from $3$ to $4$ as $4$ is already infected. A new infection event starting from $3$ is inserted into the queue.\textbf{ (d)} Node $4$ recovers, the remaining queue is shown.
	}
	\label{fig:fullpage}
\end{figure}     
	
	\subsection{Analysis}
	Our approach combines the advantages of an event-based simulation with the advantages of rejection sampling. 
	In contrast to the \emph{Optimized Gillespie Algorithm}, finding the node for the next event can be done in constant time. 
	More importantly, the number of rejection events is dramatically minimized because the queue only contains events that are realistically possible.
	Therefore, it is crucial that each node ``knows'' its own curing time and that the curing events are always generated before the infection events. 
	In contrast to traditional event-based simulation, we do not have to iterate over all neighbors of a newly infected node followed by a potentially costly reordering of the queue. 
	
	\subsubsection{Runtime} 
 	For the runtime analysis, we assume that a binary heap is used to implement the event queue and that the graph structure is implemented using a hashmap.
	Each simulation step starts by taking an element from the queue (cf. Line 7, Alg. \ref{mainproc}), which can be done in constant time. 
	Applying the change of state to a particular node has constant time complexity on average and linear time complexity (in the number of nodes) in the worst case as it is based on lookups in the hashmap. 
	
	Now consider the generation of infection events. 
	Generating a waiting time (Line 3, Alg. \ref{gen_inf}) can be done in constant time because we know the degree (and therefore the rate) of each node. 
	Likewise, sampling a random neighbor (Line 8) is constant in time (assuming the number of neighbors fits in an integer). 
	Checking for an \emph{early reject} (Line 9) can also be done in constant time because each neighbor is sampled with the same (uniform) probability and is annotated with its recovery time.
	Even though each early rejection can be computed in constant time, the number of early rejections can of course increase with the mean (and maximal) degree of the network. 
	Inserting the newly generated infection event(s) to the event queue (Line 11) has a worst-case time complexity of $\mathcal{O}(\log n)$, where $n$ is the number of elements in the heap. 
	In our case, $n$ is bounded by twice the number of infected nodes. However, we can expect constant insertion costs on average \cite{hayward1991average,porter1975random}. 
	
	\subsubsection{Correctness} 
	Here, we argue that our method generates correct sample trajectories the underlying Markov model. 
	To see this, we assume some hypothetical changes to our method that do not change the sampled trajectories but  makes it easier to reason about the correctness. 
	First, assume that we abandon \emph{early rejects} and insert all events in the event queue regardless of their possibility of success. 
	Second, assume that we change the generation of infection events such that we do not only generate the earliest attempt but all infection attempts until recovery of the node. 
	Note that we do not do this in practice, as this would lead to more rejections (less early rejections). 
	
	Similar to \cite{cota2017optimized}, we find that our algorithm is equivalent to the direct event-based implementation of the following spreading process:
	
	\begin{equation*}
	\hspace{0.3cm}
	\text{I}  \xrightarrow{\mu} \text{S}    \hspace{1.5cm}
	\text{S}+\text{I}  \xrightarrow{\lambda} \text{I}+\text{I}  \hspace{1.5cm}
	\text{I}+\text{I}  \xrightarrow{\lambda} \text{I}+\text{I}   \;.
	\end{equation*}
	
In \cite{cota2017optimized},	$\text{I}+\text{I}  \xrightarrow{\lambda} \text{I}+\text{I}$ is called a shadow process, because the application of the rule does not change the network state. 
	Hence, rejections of infections in the SIS model can be interpreted as applications of the last rule in the shadow process. 
	Note that the rate at which this rule is applied to the network is the rate of the rejection events. 
	Hence, the rate at which an infected node attacks its neighbors (no matter whether in state I or S) is exactly $\lambda k$, where $k$ is the degree of the node. 
	Our  method simulates the shadow process (which is equivalent to simulating the SIS process)  in the following way:
	For each S--I edge and I--I edge, an infection event is generated with rate $\lambda$ and inserted into the queue. 
	The decision if this event will be a real or a ``shadow infection'' is postponed until the event is actually applied.   
	This is possible because both rules have the same rate, in particular, the joint rate at which an infected $k$-degree node attacks its neighbors will always be $k \lambda$.
		
	\subsection{Generalizations}
	So far we have only considered SIS processes on static and unweighted networks. This section shorty discusses how to generalize our simulation method to SIS-type processes on temporal and weighted networks. 
	
	\subsubsection{General Epidemic Models}
	A key ingredient to our algorithm is the early rejection of infection events. This is possible because we can compute the time of a node's curing already when the node gets infected.   
	In particular, we exploit that there is only one way to leave state I, that is, by the application of a node-based rule. This gives us a guarantee about the remaining time in state I.
	Other epidemic models have a similar structure. For instance, consider the Susceptible-Infected-Recovered (SIR, sometimes also referred to as SIRS) model, where infected nodes first become recovered (immune), before entering state I again:
	
	\begin{equation*}
	\hspace{0.3cm}
	\text{S}+\text{I}  \xrightarrow{\lambda} \text{I}+\text{I}  \hspace{1.5cm}
	\text{I}  \xrightarrow{\mu_1} \text{R}    \hspace{1.5cm} 
	\text{R}  \xrightarrow{\mu_2} \text{S}    \;,
	\end{equation*}
	
We also consider the competing pathogens model \cite{masuda2006multi}, where two infectious diseases, denoted by I and J, compete over the susceptible nodes:
	
	\begin{equation*}
	\hspace{0.3cm}
	\text{S}+\text{I}  \xrightarrow{\lambda_1} \text{I}+\text{I}  \hspace{1.2cm}
	\text{S}+\text{J}  \xrightarrow{\lambda_2} \text{J}+\text{J}  \hspace{1.2cm}
	\text{I}  \xrightarrow{\mu_1} \text{S}    \hspace{1.2cm}  
	\text{J}  \xrightarrow{\mu_2} \text{S}   \;.
	\end{equation*}
	
	In both cases, we can exploit that certain states (I, J, R) can only be left under node-based rules and thus their residence time is independent of their neighborhood. 
	This makes it simple to annotate each node in any of these states with their exact residence time and perform early rejections accordingly.
	
	If we do not have these guarantees, early rejection cannot be applied. For instance in the (fictional) system: 
	
	\begin{equation*}
	\hspace{0.9cm}
	\text{S}+\text{I}  \xrightarrow{\lambda_1} \text{I}+\text{I}  \hspace{1.5cm}
	\text{I}+\text{I}  \xrightarrow{\lambda_2} \text{I}+\text{S}  \;.
	\end{equation*}
	
	It is likely that  our method will  still perform better than the traditional event-based approach, however, the number of rejection events might significantly decrease its performance.
	
	\subsubsection{Weighted Networks}
	In weighted networks, each edge $e \in \mathcal{E}$ is associated with a positive real-valued weight $w(e) \in \mathbb{R}_{> 0}$. 
	Each edge-based rule of the form
		\begin{center}$
 \text{A}+\text{C} \xrightarrow{\lambda} \text{B}+\text{C}$
 	\end{center}
	  fires on this particular edge with rate $w(e) \cdot \lambda$. 
	  Hence, unweighted networks can be seen as weighted networks  with all weights being $1$.
	Applying  our method to weighted networks is simple:  Let $n$ be a  node. 
	During the generation of infection events, instead of sampling the waiting time with rate $\lambda k$, we now use rate $\lambda \sum_{n' \in N(n)} w(n,n')$, where $N(n)$ is the set of neighbors of $n$.  
	Moreover, instead of choosing a neighbor that will be attacked with uniform probability, we choose them with a probability proportionally to their edge weight. 
	This can either be done by rejection sampling or in $\mathcal{O}(\log (k))$ time complexity, where $k$ is the degree of $n$. 
	
	\subsubsection{Temporal Networks}
	Temporal (time-varying, adaptive, dynamic) networks are an intriguing generalization of static networks which generally complicates the analysis of  their spreading behavior
	\cite{vestergaard2015temporal,masuda2017temporal,holme2012temporal,holme2015modern}.
	Generalizing the Gillespie algorithm for Markovian epidemic-type processes is far from trivial \cite{vestergaard2015temporal}. 
	
	In order to keep our model as general as possible, we assume here that an external process governs the temporal changes in the network.
	This process runs simultaneously to our simulation and might or might not depend on the current network state. It changes the current graph by adding or removing edges, one edge at a time.  
	For instance, after processing one event, the external process could add or remove an arbitrary (but finite) number of edges at specific time points until the time of the next event is reached. 
	It is simple to integrate this into our simulation. 
	
	Given that the external process removes an edge, we can simply update the neighbor list and the degrees in our graph. 
	For each infection event that reaches the top of the queue, we first check if the corresponding edge is still present.
	If not, we reject the event. This is possible because removing events only decreases infection rates which we can correct by using rejections. 
	When an edge is added to the graph and at least one corresponding node is infected, 
	the infection rate increases. 
	Thus, it is not sufficient to only update the graph, we also generate an infection event which accounts for the new edge. 
	In order to minimize the number of generated events, we change the algorithm such that each infected node is annotated with the time point of its subsequent infection attempt. 
	Consider now an infected node. When it obtains a new edge, we generate an exponentially distributed waiting time with rate $\lambda$ modeling the infection attempt through this specific link.
	We only generate a new event if this  time point lies before the time point of the subsequent infection attempt of the node.
	In that case, we also remove the old event associated with this node from the queue. 
	
	Since most changes in the graph do not require changes  of the event queue (and those 
	that do only cause   two operations at maximum), we expect our method to handle temporal networks with a reasonably high number of graph updates very efficiently.
 	In the case that an extremely large number of edges in the graph change at once, we can always decide to iterate over the whole network and newly initialize the event queue.

	\section{Case Studies\label{casestudies}}
	We demonstrate the effectiveness our approach on three classical epidemic-type processes.  
	We compare the performance of our method with the \emph{Standard Gillespie Algorithm} (GA) and the \emph{Optimized Gillespie Algorithm} (OGA) for different network sizes. 
	We use synthetically generated networks following the configuration model \cite{fosdick2018configuring} with a truncated power-law degree distribution, that is $P(k) \propto k^{-\gamma}$ for $3  \leq k  \leq 1000$. 
	We compare the performance on degree distributions with $\gamma \in \{2,3\}$.
	This yields a mean degree around $30$ ($\gamma=2$) and $10$ ($\gamma=3$). 
	We use models from literature but adapt rate parameters freely to cause interesting dynamics.
	
	In particular, we report how the number of nodes in a network is related to the CPU time of a single step. 
	This is more informative than using the total runtime of a simulation because the number of steps obviously increases with the number of nodes when the time horizon is fixed. 
	The CPU time per step is defined as the total runtime of the simulation divided by the number of steps, only counting the steps that actually change the network state (i.e.,\;excluding rejections). 
	We do not count rejection events, because that would give an unfair advantage to the rejection based approach. The evaluation was performed on a 2017 MacBook Pro with a 3.1 GHz Intel Core i5 CPU and 16 GB of RAM. 
	
	A comparison with other tools is difficult as they typically report neither CPU time nor derive asymptotic guarantees. 
	Comparison with other tools is difficult as they typically report neither CPU time nor derive asymptotic guarantees, one positive exception is \cite{sahneh2017gemfsim} who also implement event-based simulation, but one suited for mean-field type approximations of networks.
	
	\subsection{SIS Model}
	For the SIS model we used rate parameters of $(\mu, \lambda) = (1.0, 0.6)$ and an initial distribution of $95\%$ susceptible nodes and $5\%$ infected nodes.
	CPU times are reported in Fig.~\ref{fig:resultfileplotfinalSIS}a, where ``reject'' refers to our
	rejection-based algorithm (as described in Section~\ref{ourmethod}).
	For a sample trajectory, we plot the fraction of nodes in each state w.r.t.\;time (Fig. \ref{fig:resultfileplotfinalSIS}b). 
	To have a comparison with OGA we used the official Fortran-implementation in  \cite{cota2017optimized}  and estimated the average CPU time per step based on the absolute runtime. 
	Note that the comparison is not perfectly fair due to implementation differences and additional input/output of the OGA code.
	It is not surprising that the OGA performs comparably bad, as the method is suited for simulations close to the epidemic threshold. 
	Moreover, our maximal degree is very large, which negatively effects the performance of the OGA.

	\begin{figure}[t!]
		\centering
		\subfloat[]{{ \includegraphics[width=0.6 \linewidth]{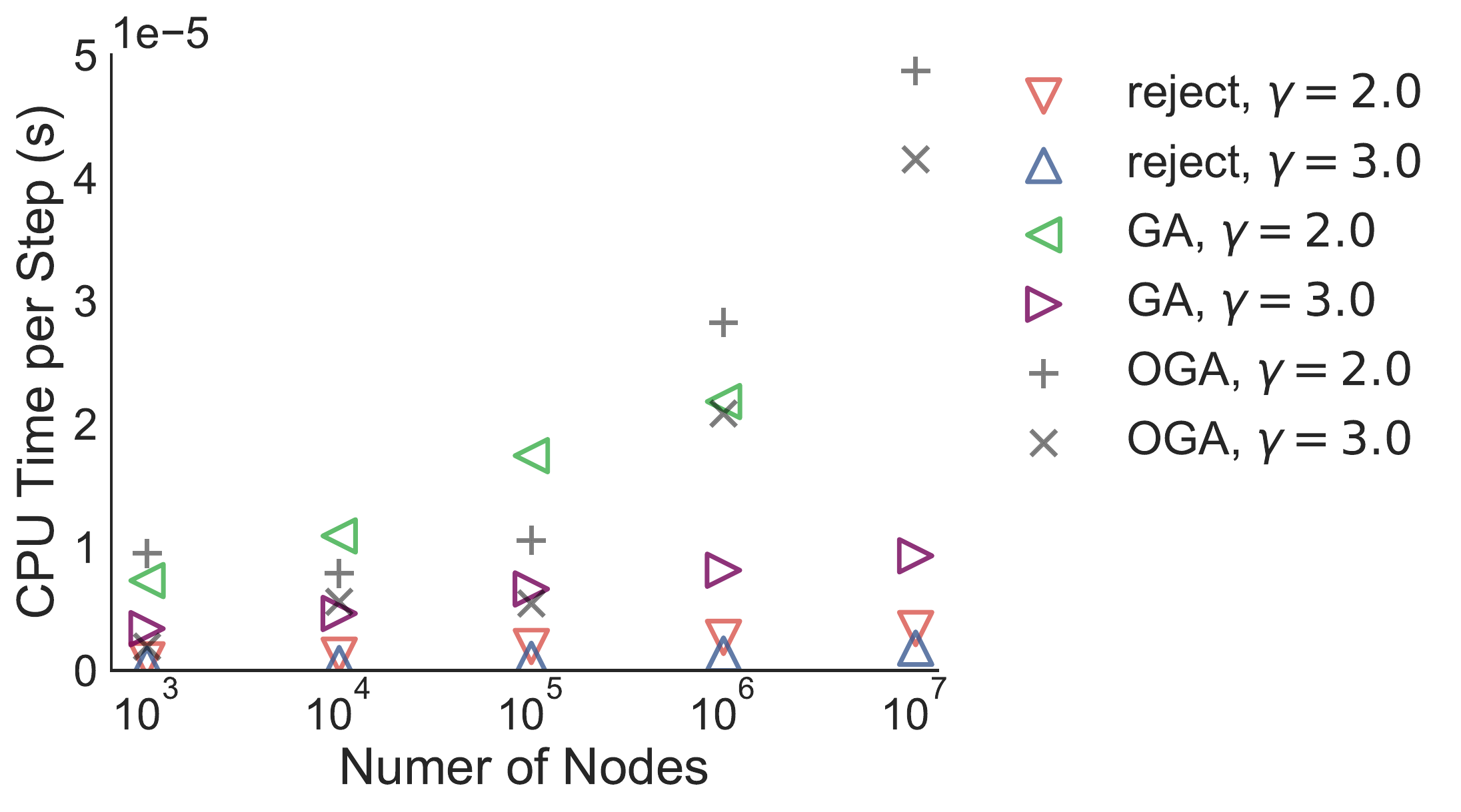} }}%
		  \subfloat[]{{\raisebox{0.6cm}{\includegraphics[width=0.35\linewidth]{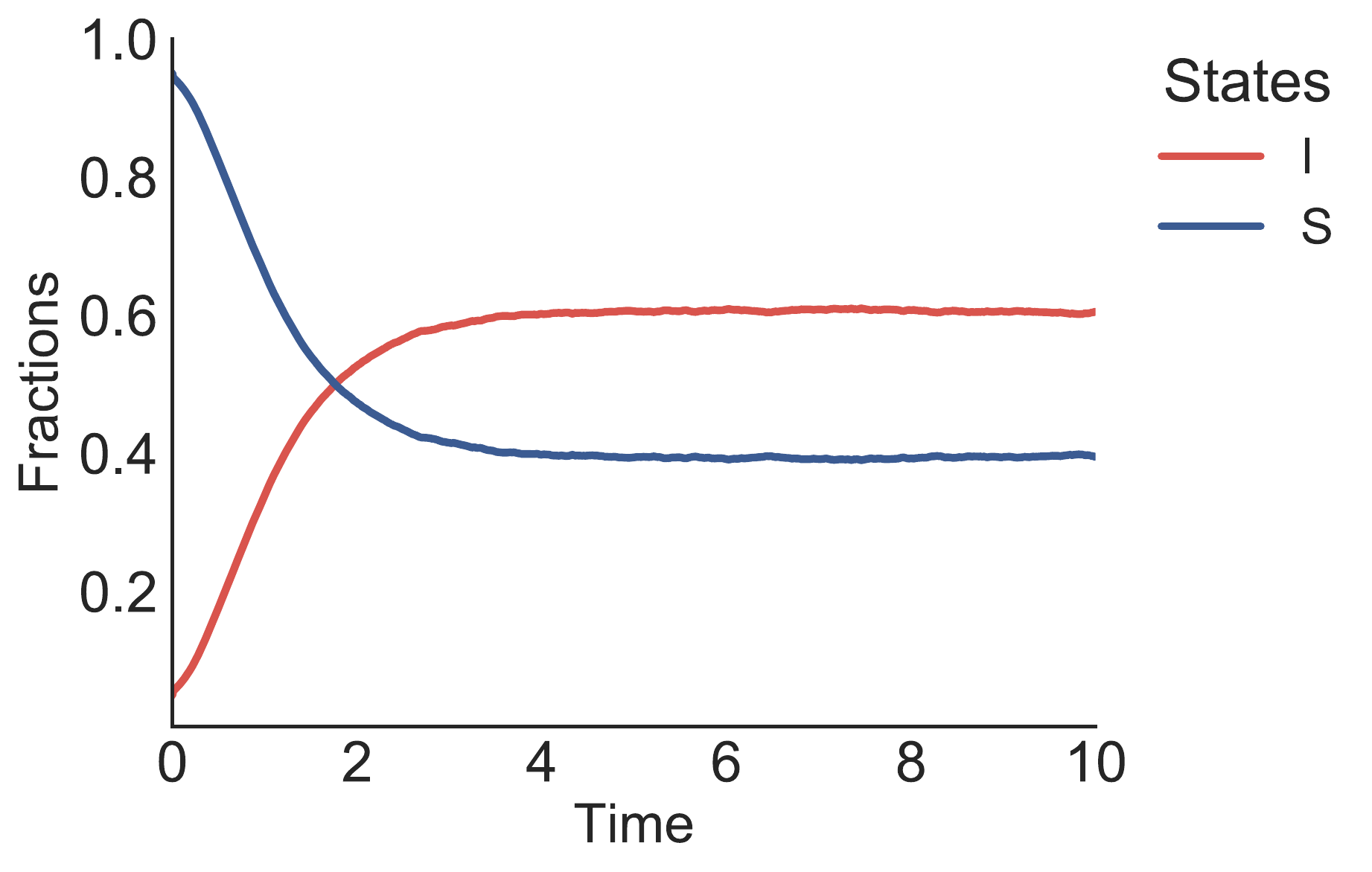} }}}%

		\caption{SIS model (a): Average CPU time for a single step (i.e., change of network state) for different networks. 
		The GA method run out of memory for $\gamma=2.0$, $|N|=10^7$.
		(b):  Sample dynamics for a network with $\gamma=3.0$ and $10^5$ nodes. 
		\label{fig:resultfileplotfinalSIS}} 
	\end{figure}

	\subsection{SIR Model}
	Next, we considered the SIR model, which has more complex dynamics.
	We used rate parameters of $(\mu_1, \mu_2, \lambda) = (1.1, 0.3, 0.6)$ and an initial distribution of $96\%$ susceptible nodes and $2\%$ infected and recovered nodes, respectively. 
	Similar as above, CPU times and example dynamics are reported in Fig.~\ref{fig:resultfileplotfinalSIR}. We see that runtime behavior is almost the same as in the SIS model.
	Note that an implementation of the OGA was only available for the SIS model and the comparison is therefore
	not available for other models. 
	Due to the high number of rejection steps, we expect a similar performance difference.
  	
	\begin{figure}[t!]
		\centering
		\subfloat[]{{ \includegraphics[width=0.6 \linewidth]{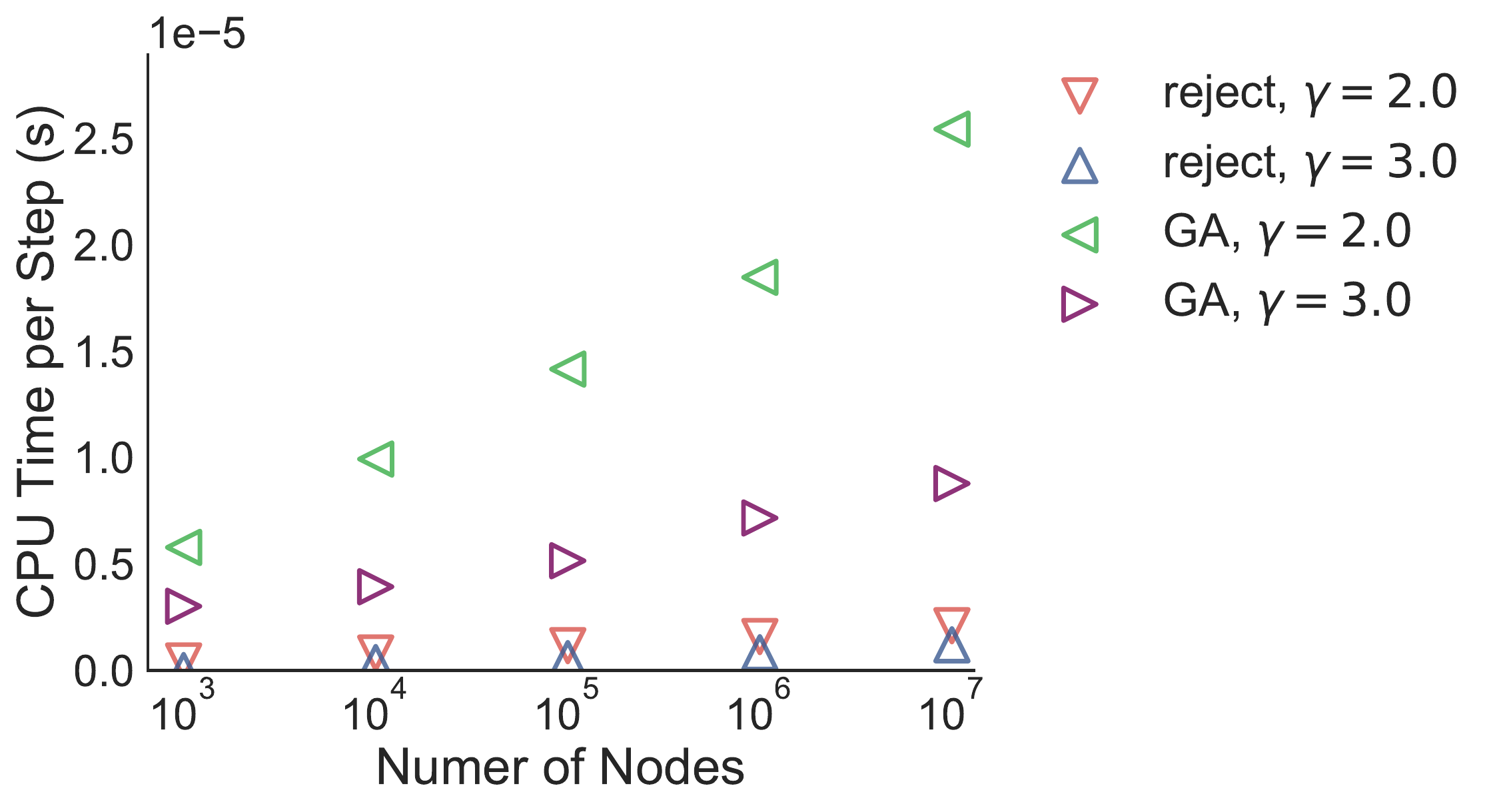} }}%
		  \subfloat[]{{\raisebox{0.6cm}{\includegraphics[width=0.35\linewidth]{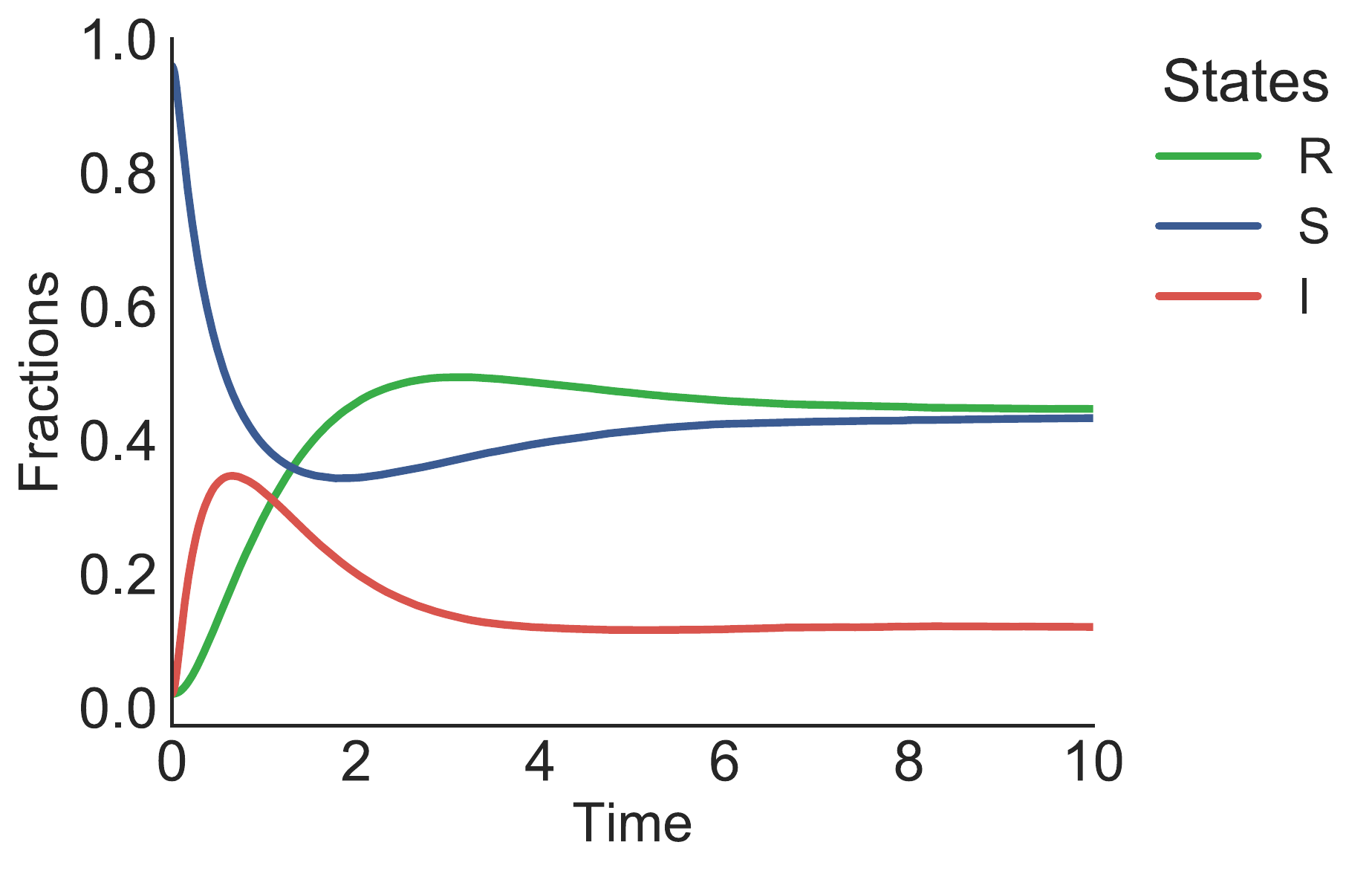} }}}%

		\caption{SIR model (a): Average CPU time for a single step (i.e., change of network state) for different networks.  (b):  Sample dynamics for a network with $\gamma=2.0$ and $10^5$ nodes.\label{fig:resultfileplotfinalSIR}} 
		
	\end{figure}
	
	\subsection{Competing Pathogens Model}
	Finally, we considered the Competing Pathogens model. 
	We used rate parameters of $( \lambda_1, \lambda_2, \mu_1, \mu_2) = (0.6, 0.63, 0.6, 0.7)$ and an initial distribution of $96\%$ susceptible nodes and $2\%$ infected nodes for both pathogens (denoted by I, J), respectively. 
	CPU times and network dynamics are reported in Fig.~\ref{fig:resultfileplotfinalSIJS}.
	The model is interesting because we see that in the beginning J dominates I due to its higher infection rate. However, nodes infected with pathogen J recover faster than those infected with I. This gives the I pathogen the advantage that infected nodes have more time to attack their neighbors. In the limit, I takes over and J dies out. 
For	this model  stochastic noise has a significant influence on the macroscopic dynamics.
Therefore,  we also reported the standard deviation of the fractions (cf. Fig.~\ref{fig:resultfileplotfinalSIJS}).
Note that the fraction of susceptible nodes is almost deterministic. 
	Performance-wise our rejection method performs slightly worse than in the previous models (w.r.t. the baseline).
We believe that this is due to the large number of infection events and rejections. 
	
	\begin{figure}[t!]
		\centering
		\subfloat[]{{ \includegraphics[width=0.6 \linewidth]{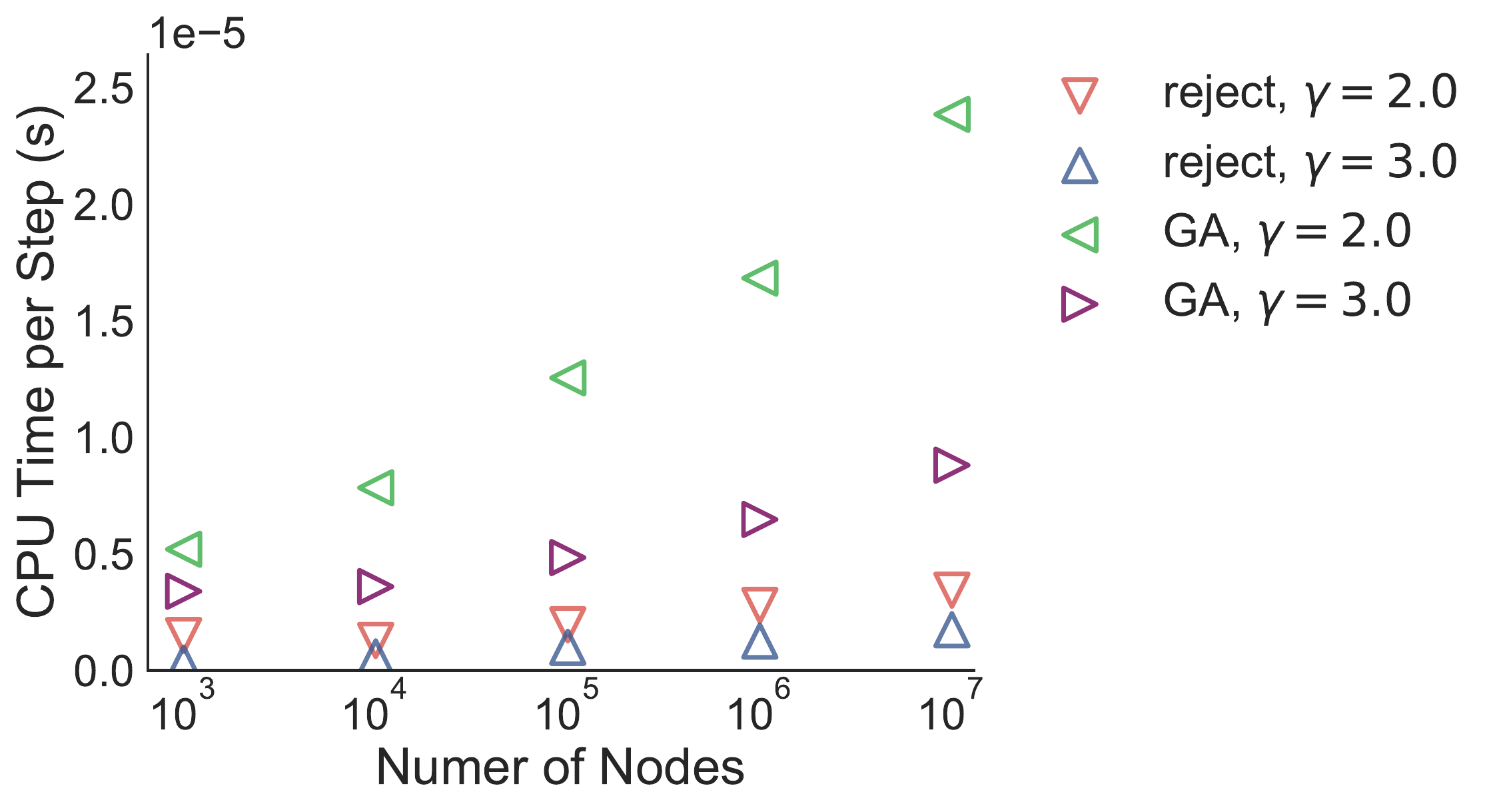} }}%
		  \subfloat[]{{\raisebox{0.6cm}{\includegraphics[width=0.35\linewidth]{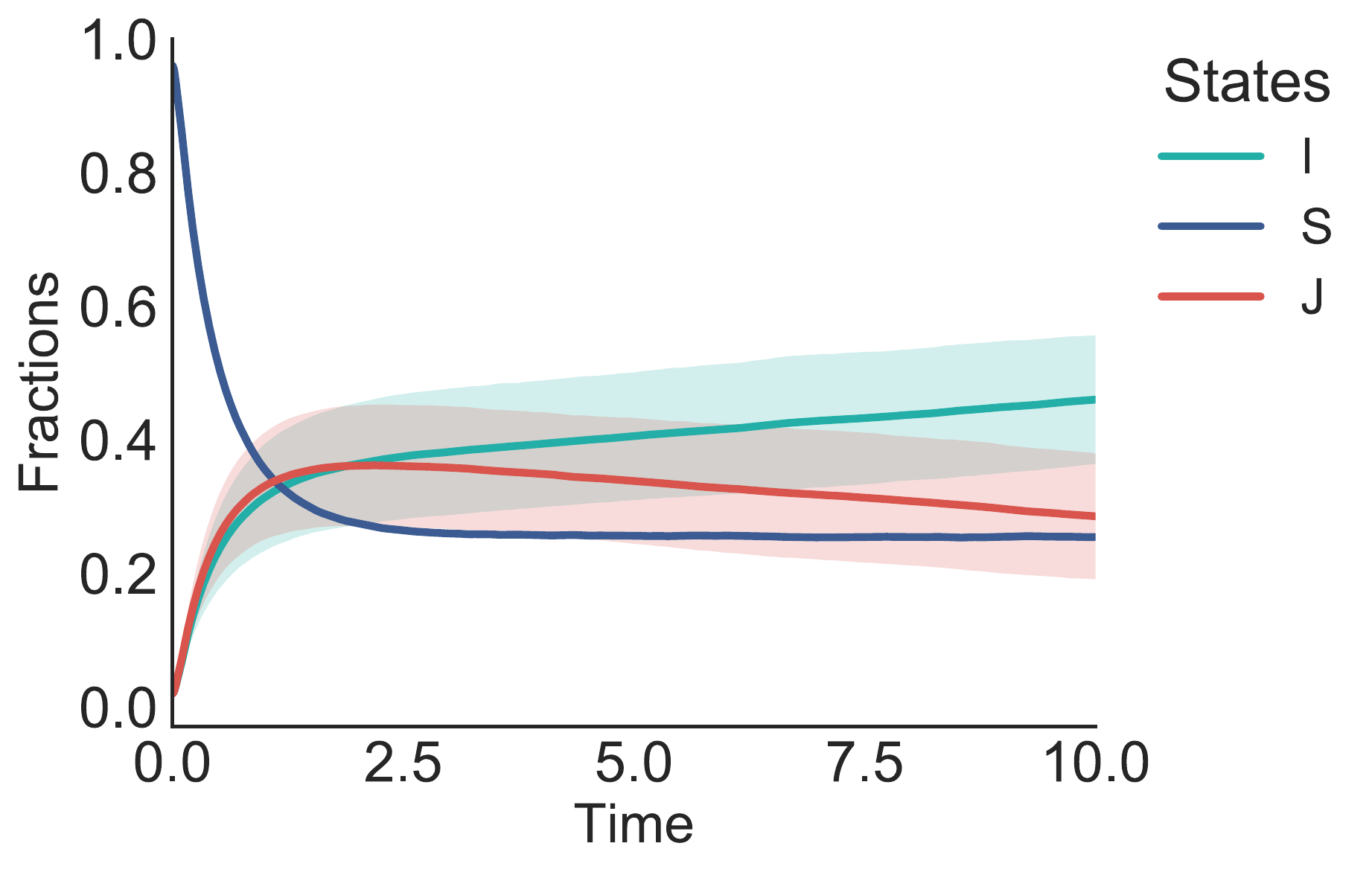} }}}%

		\caption{Competing pathogens model (a): Average CPU time for a single step (i.e., change of network state) for different networks.  (b):  Mean fractions and standard deviations of   a network with $\gamma=2.0$ and $10^4$ nodes.\label{fig:resultfileplotfinalSIJS}} 
		
	\end{figure}

	\section{Conclusions\label{conclusions}}
	In this paper, we presented a novel rejection algorithm for the simulation of epidemic-type processes. We combined the advantages of rejection sampling and event-driven simulation. In particular, we exploited that nodes can only leave certain states using node-based rules, which made it possible to precompute their residence times, which then again allowed us to perform early rejection of certain events. 
	
	Our numerical results show that our method outperforms previous approaches especially in networks which are not close the epidemic threshold. 
	significantly better than previous ones. 
	In particular, the speed-up increases as the maximal degree of the network increases.
	
	As future work, we plan to extend the method to compartment models with arbitrary rules, including an automated decision for which states early rejections can be computed and are useful. 
	

	\subsubsection{Acknowledgments}
    We thank Michael Backenköhler for his comments on the manuscript.

	\newpage
	\bibliographystyle{unsrt}

\end{document}